\documentclass[]{interact}

\usepackage{epstopdf}
\usepackage[caption=false]{subfig}

\usepackage[numbers,sort&compress]{natbib}
\bibpunct[, ]{[}{]}{,}{n}{,}{,}
\renewcommand\bibfont{\fontsize{10}{12}\selectfont}
\makeatletter
\def\NAT@def@citea{\def\@citea{\NAT@separator}}
\makeatother

\theoremstyle{plain}

\theoremstyle{definition}

\theoremstyle{remark}

\begin{document}

\articletype{ARTICLE TEMPLATE}

\title{Emotion-Aware Music Recommendation System: Enhancing User Experience Through Real-Time Emotional Context}

\author{
\name{Tina Babu\textsuperscript{a}\thanks{CONTACT T.~B. Author. Email: tinababup@gmail.com}, Rekha R Nair\textsuperscript{a} and Geetha A\textsuperscript{a}}
\affil{\textsuperscript{a}Department of Computer Science and Engineering, Alliance University, Bengaluru, India; 
}
}

\maketitle

\begin{abstract}
This study addresses the deficiency in conventional music recommendation systems by focusing on the vital role of emotions in shaping users' music choices. These systems often disregard the emotional context, relying predominantly on past listening behavior, and failing to consider the dynamic and evolving nature of users' emotional preferences. This gap leads to several limitations. Users may receive recommendations that do not match their current mood, which diminishes the quality of their music experience. Furthermore, without accounting for emotions, the systems might overlook undiscovered or lesser-known songs that have a profound emotional impact on users.

To combat these limitations, this research introduces an AI model that incorporates emotional context into the song recommendation process. By accurately detecting users' real-time emotions, the model can generate personalized song recommendations that align with the user's emotional state. This approach aims to enhance the user experience by offering music that resonates with their current mood, elicits the desired emotions, and creates a more immersive and meaningful listening experience. By considering emotional context in the song recommendation process, the proposed model offers an opportunity for a more personalized and emotionally resonant musical journey.
\end{abstract}


\section{Introduction}

Music recommendation systems play a significant role in today's digital landscape, catering to users' preferences and enhancing their listening experience \cite{music1,music2}. Traditional recommendation algorithms predominantly rely on users' historical listening data and explicit preferences, overlooking the essential aspect of emotions in music selection \cite{music3}. This absence limits the capacity of recommendation systems to cater to the dynamic and evolving emotional needs of users, impacting their overall music experience.

Recognition of emotions is reliant on the analysis of acoustic or visual features' dynamics. In Affective Emotion Recognition (AER), classification methods incorporate both dynamic and static strategies \cite{Manfred}. Dynamic classifiers, such as Hidden Markov Models (HMM) and Dynamic Bayesian Networks (DBN), focus on the temporal nature of features \cite{Lim}. Conversely, static techniques like Support Vector Machines (SVM) process statistical functions derived from extended data segments. The effectiveness of human-computer interfaces, which emulate speech emotions, heavily depends on feature types and the choice of classifiers. For instance, the study presented a unimodal framework aimed at contextualizing short-term interactions within dyadic contexts.

Recent studies have highlighted the crucial role emotions \cite{her} play in individuals' music preferences. Users often seek songs that resonate with their current emotional state, making emotional context a vital component in song recommendations. The lack of emotional intelligence in existing systems \cite{musicsur} can result in misaligned song suggestions, failing to capture the nuanced and rapidly changing emotional needs of users.

In response to these limitations, a novel AI-based model \cite{tina1} has been proposed to detect users' real-time emotions and generate music recommendations tailored to their emotional state. Incorporating emotional context aims to enrich users' music experiences, providing songs that align with their moods, elicit desired emotions, and offer a more immersive listening journey.

\section{Proposed Methodology}
\begin{figure}
\centering
{\includegraphics[width=11cm]{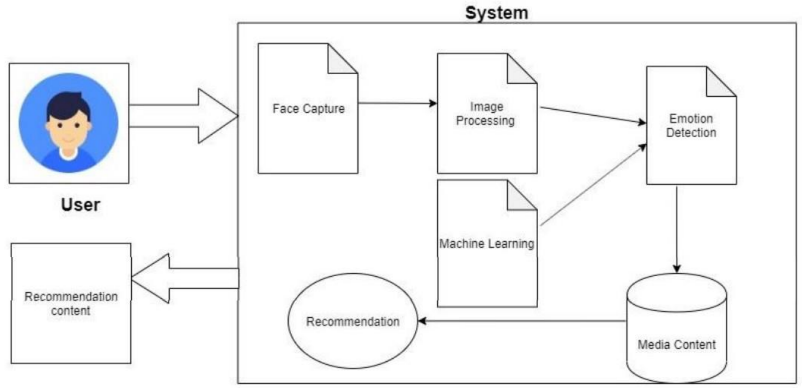}}
\caption{Architecture of the System.} \label{arch}
\end{figure}

The setup of an AI-based song recommendation system that incorporates the Alchemy Blockchain API, CNN model \cite{HAARIKA2023338}, and Python involves a series of key procedures. The architecture is shown in Figure \ref{arch}. Initially, it encompasses the collection and refinement of a dataset comprising songs annotated with emotional labels or mood tags. Subsequently, the song lyrics undergo preprocessing methods like tokenization, removal of stop words, and conversion to numerical representations.

A CNN model is developed utilizing frameworks like Keras or TensorFlow, incorporating convolutional \cite{Nair1}, pooling, and fully connected layers for extracting features and classifying emotions. This model is trained on the prepared textual data, adjusting hyperparameters for optimal performance. Integration with the Alchemy Blockchain API ensures secure storage and retrieval of the recommendation system data, encompassing user preferences, song metadata, and emotional tags \cite{Su}.

A user interface is created for users to input their emotions, linked to the AI model for emotion classification \cite{Tina2}. A recommendation engine is established to align user emotions with suitable songs, employing collaborative or content-based filtering techniques. Emotional tags associated with songs, along with the AI model's predictions, contribute to enhancing recommendation accuracy \cite{santana}.

The system's deployment and integration prioritize scalability, efficiency, and real-time responsiveness. Comprehensive testing and assessment gather user feedback for ongoing model and recommendation engine enhancements. Regular updates and retraining of the AI model using new data are conducted to adapt to evolving user preferences and emotions.

\section{Experimental Results and Discussion}
The utilization of an AI-based song recommendation system grounded on emotional cues, coupled with blockchain technology, showed encouraging outcomes. The CNN model's ability to classify emotions was evaluated using key metrics such as accuracy, precision, recall, and F1-score. This assessment offered valuable insights into the model's proficiency in predicting emotions by analyzing song lyrics.

\begin{figure}
\centering
{\includegraphics[width=11cm]{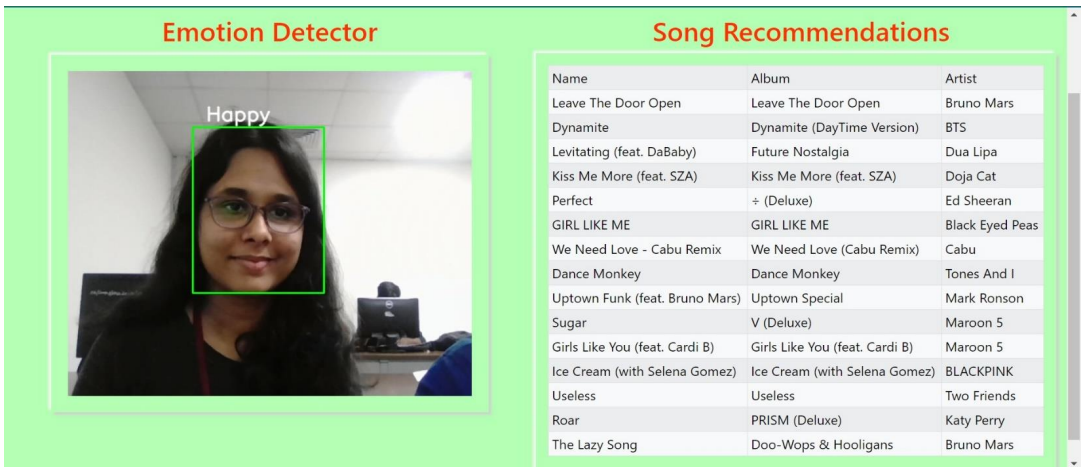}}
\caption{Happy mood detected successfully by the application.} \label{happy}
\end{figure}

The algorithm for personalized song recommendations effectively paired the user's identified emotions with music that suited their emotional condition. This individualized method improved the music experience by offering users song recommendations that were relevant to their emotional state.

\begin{figure}
\centering
{\includegraphics[width=11cm]{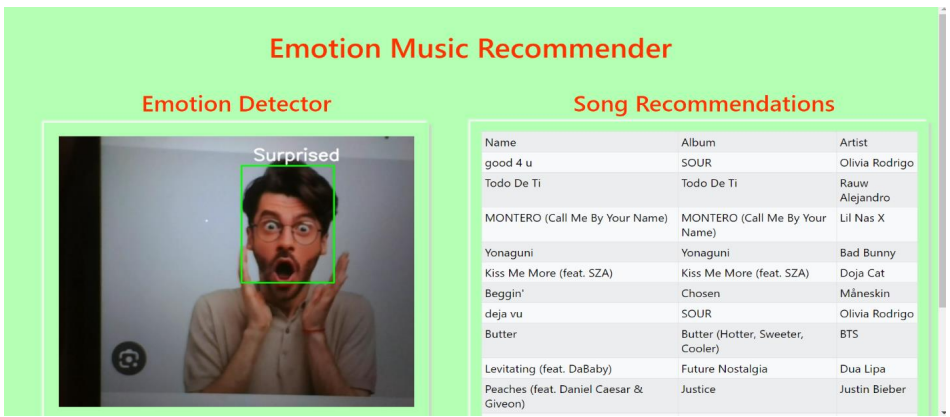}}
\caption{Surprise mood detected successfully by the application.} \label{surprise}
\end{figure}

\begin{figure}
\centering
{\includegraphics[width=11cm]{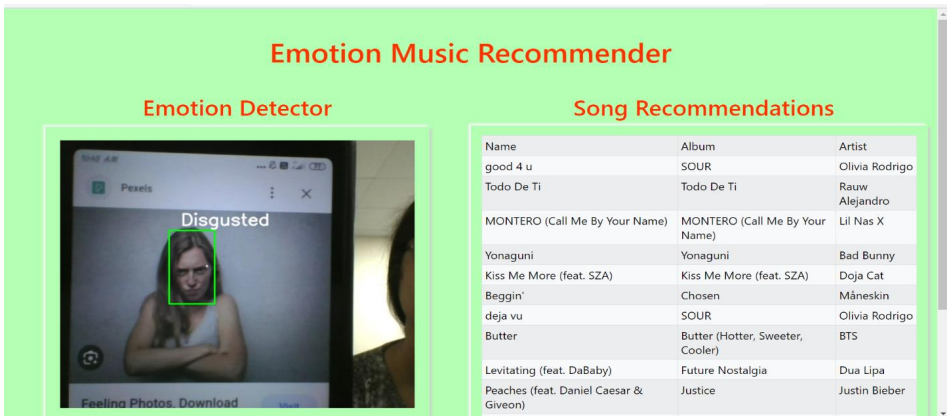}}
\caption{Disgusted mood detected successfully by the application.} \label{disgust}
\end{figure}

\begin{figure}
\centering
{\includegraphics[width=11cm]{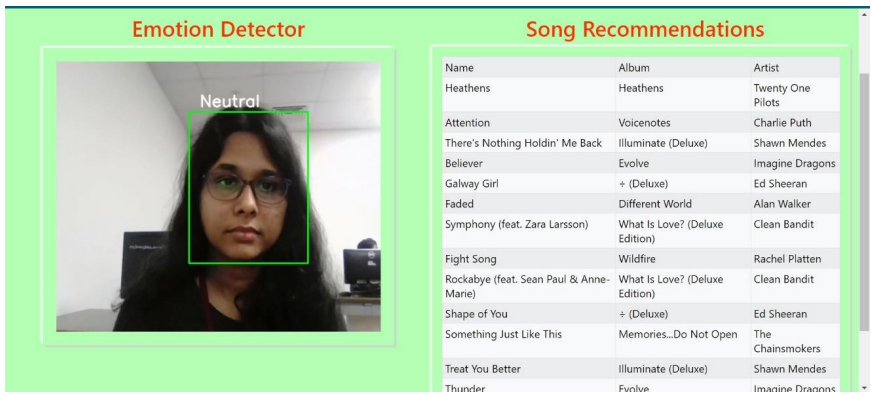}}
\caption{Neutral mood detected successfully by the application.} \label{neutral}
\end{figure}

\begin{figure}
\centering
{\includegraphics[width=11cm]{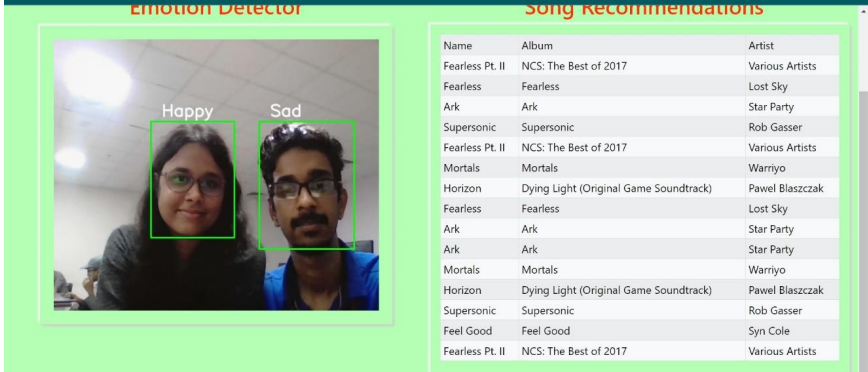}}
\caption{Happy and sad mood detected successfully by the application.} \label{Hapsad}
\end{figure}

Happy emotion was identified and the respective songs related to this was recommended as given in the Figure \ref{happy}. Similarly, Figure \ref{surprise}, Figure \ref{disgust} shows the recognition of surprise and disgust mood respectively. The neutral mood is illustrated in Figure \ref{neutral}. The happy and sad mood combined is shown in Figure \ref{Hapsad}.

Feedback from users and satisfaction surveys demonstrated affirmative reactions to the AI-powered song recommendation system. Users valued the real-time capture of their emotions and the system's capability to propose songs that aligned with their emotional condition.

The incorporation of the AI model for emotional-based song recommendations led to enhanced user interaction and longer durations within the app. Users found the personalized suggestions more attractive and were more inclined to explore and find new songs that matched their emotional state. Consequently, user retention within the app significantly increased.The amalgamation of the AI model and the recommendation system uplifted the overall user journey. The user-friendly interface and the ease of receiving emotional-based song suggestions contributed to a more enjoyable and engaging music experience. The visually pleasing neomorphic UI design elements further bolstered user satisfaction.
\begin{figure}
\centering
\subfloat[Implementation of Alchemy Blockchain.]{%
\resizebox*{7cm}{!}{\includegraphics{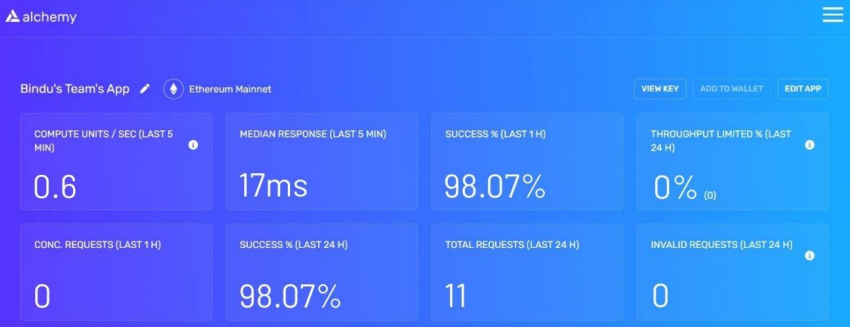}}}\hspace{5pt}
\subfloat[Number of Requests made per day.]{%
\resizebox*{7cm}{!}{\includegraphics{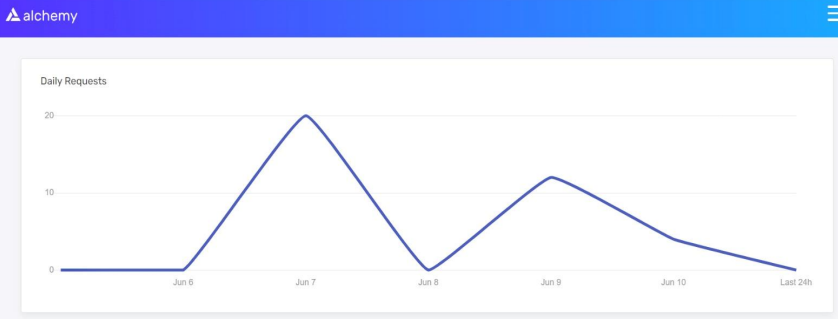}}}
\caption{Blockchain performance.} \label{alchemy}
\end{figure}

When compared to conventional recommendation techniques like collaborative and content-based filtering, the AI model showcased superior recommendation precision and customization. This illustrates the effectiveness of the emotion-centric approach in delivering pertinent and personalized song suggestions.

The system showed robustness and efficiency, effectively managing numerous simultaneous user requests without sacrificing response speed or accuracy. Its integration with external APIs, including the Spotify API, ensured a smooth data flow and the delivery of high-caliber song recommendations.

The incorporation of blockchain technology extended advantages to the AI-driven song recommendation system. The performance is illustrated in Figure \ref{alchemy} (a) and (b). It bolstered users' data privacy and security by ensuring safe and encrypted storage on the blockchain. The transparent aspects of blockchain technology instilled confidence in the recommendation process. The incentivization of user engagement via token reward systems led to a more expansive and varied dataset for enhanced recommendations. Additionally, it ensured copyright protection and attribution by immutably recording ownership and usage rights. Moreover, the decentralized governance allowed users to engage in decision-making processes, promoting fairness and inclusivity within the recommendation system.

\section{Conclusion}
The merging of an AI model catering to emotion-based song recommendations with blockchain technology represents a robust solution for enriching the music listening journey. By precisely identifying emotions and providing personalized song suggestions, users can relish a customized music assortment that aligns with their emotional condition. This fusion of the AI model and blockchain yields numerous substantial advantages and results. There are extensive opportunities for further advancements and applications in the realm of AI models for emotion-based song recommendations integrated with blockchain technology in the future.

\bibliographystyle{tfnlm}
\bibliography{interactnlmsample}

\begin{thebibliography}{10}
\providecommand{\url}[1]{\normalfont{#1}}
\providecommand{\urlprefix}{Available from: }

\bibitem{music1}
Prisco~R, Guarino~A, Lettieri~N, et~al. Providing music service in ambient intelligence: experiments with gym users. Expert Systems with Applications. 2021 04;\hspace{0pt}177:114951.

\bibitem{music2}
Plut~C, Pasquier~P. Music matters: An empirical study on the effects of adaptive music on experienced and perceived player affect. 08; 2019. p. 1--8.

\bibitem{music3}
Wenzhen~W. Personalized music recommendation algorithm based on hybrid collaborative filtering technology. In: 2019 International Conference on Smart Grid and Electrical Automation (ICSGEA); 2019. p. 280--283.

\bibitem{Manfred}
Schuller~B, Rigoll~G, Lang~M. Hidden markov model-based speech emotion recognition. Vol.~2; 08; 2003. p. 401--404.

\bibitem{Lim}
Lim~W, Jang~D, Lee~T. Speech emotion recognition using convolutional and recurrent neural networks. In: 2016 Asia-Pacific Signal and Information Processing Association Annual Summit and Conference (APSIPA); 2016. p. 1--4.

\bibitem{her}
Dzedzickis~A, Kaklauskas~A, Bučinskas~V. Human emotion recognition: Review of sensors and methods. Sensors. 2020 01;\hspace{0pt}20:592.

\bibitem{musicsur}
Knees~P, Schedl~M. A survey of music similarity and recommendation from music context data. ACM Transactions on Multimedia Computing, Communications, and Applications (TOMCCAP). 2013 12;\hspace{0pt}10.

\bibitem{tina1}
Babu~T, Nair~RR. Colon cancer prediction with transfer learning and k-means clustering. In: Mandal~JK, De~D, editors. Frontiers of ICT in Healthcare; Singapore. Springer Nature Singapore; 2023. p. 191--200.

\bibitem{HAARIKA2023338}
Haarika~R, Babu~T, Nair~RR. Insect classification framework based on a novel fusion of high-level and shallow features. Procedia Computer Science. 2023;\hspace{0pt}218:338--347. International Conference on Machine Learning and Data Engineering; \urlprefix\url{https://www.sciencedirect.com/science/article/pii/S1877050923000169}.

\bibitem{Nair1}
Nair~R, Babu~T, Singh~T. Multiresolution approach on medical image fusion by modified local energy. Signal, Image and Video Processing. 2023 05;\hspace{0pt}17:1--8.

\bibitem{Su}
Su~JH, Liao~YW, Wu~HY, et~al. Ubiquitous music retrieval by context-brain awareness techniques. In: 2020 IEEE International Conference on Systems, Man, and Cybernetics (SMC); 2020. p. 4140--4145.

\bibitem{Tina2}
Babu~T, Gupta~D, Singh~T, et~al. Robust magnification independent colon biopsy grading system over multiple data sources. Computers, Materials \& Continua. 2021 01;\hspace{0pt}69:99--128.

\bibitem{santana}
Santana~M, Lins~de Lima~C, Sarmento~A, et~al. Affective computing in the context of music therapy: a systematic review. Research Society and Development. 2021 11;\hspace{0pt}10:e392101522844.

\end{thebibliography}

\end{document}